\documentclass[11pt,twoside]{article}
\usepackage{./asp2010}

\newcommand{\Msun}{\mbox{\,M}_{\odot}}

\newcommand{\grad}{^{\mbox{\small o}}}
\newcommand{\Myr}{\mbox{\,Myr}}
\newcommand{\dif}{\mathrm{d}}
\resetcounters

\begin{document}

\title{Analyzing spiral structure in a galactic disk with a gaseous component}
\author{Mata-Ch\'avez, M. Dolores,$^1$ G\'omez Gilberto C.,$^1$ and Puerari, Iv\^anio$^2$}
\affil{$^1$ Centro de Radioastronom\'ia y Astrof\'isica, UNAM}
\affil{$^2$ Instituto Nacional de Astrof\'isica \'Optica y Electr\'onica}

\begin{abstract}
Using GADGET2 \citep{Springel2001}, we performed an SPH+N-body simulation of a galactic disk with stellar and gas particles. This simulation allows to compare the spiral structure in the different disk components.  Also, we performed a simulation without gaseous component to explore the effects of the gas in the spiral pattern of the stars.
\end{abstract}

\section{Model}
 The model consists in a disk with $12\times10^{6}$ gas and stellar particles, $9.8\times10^{8} \Msun$ of gas and $3.49\times10^{10} \Msun$ of stars. Both components are initially distributed axisymmetrically with an exponential profile, in rotational equilibrium with a potential similar to that described in \cite{Allen&Santillan1991}, which includes  halo, bulge and disk components.                 
  The gaseous component includes a cooling equation (\citealt {Koyama2000}, as corrected in \citealt{Vazquez-Semadeni2007}).
A second simulation was also performed with the same conditions but without the gaseous component. 

\section{Analysis}

Using Fourier analysis,  we obtained the amplitude of the spiral modes in the evolved disk \cite{Douglas2010}.
\begin{equation}
A(m,p) =\frac{1}{D} \int_{-\pi}^{+\pi} \int_{r_{min}}^{r{ma}} \Sigma(u,\phi) \exp\left[-i(m\phi + pu\right)] \dif\phi \dif u,
\end{equation}

\begin{equation}
D = \int_{-\pi} ^{+\pi} \int_{r_{min}} ^{r_{max}}\Sigma(u,\theta) \dif u \dif\phi,
\end{equation}

\noindent where $\Sigma$ is the surface density, $u=\log r$, $m$ represents the number of arms and $p$ is associated with the pitch angle, $\alpha = \arctan(-m/p)$ 

\begin{figure*}
 \begin{center}
  \includegraphics[width=36mm]{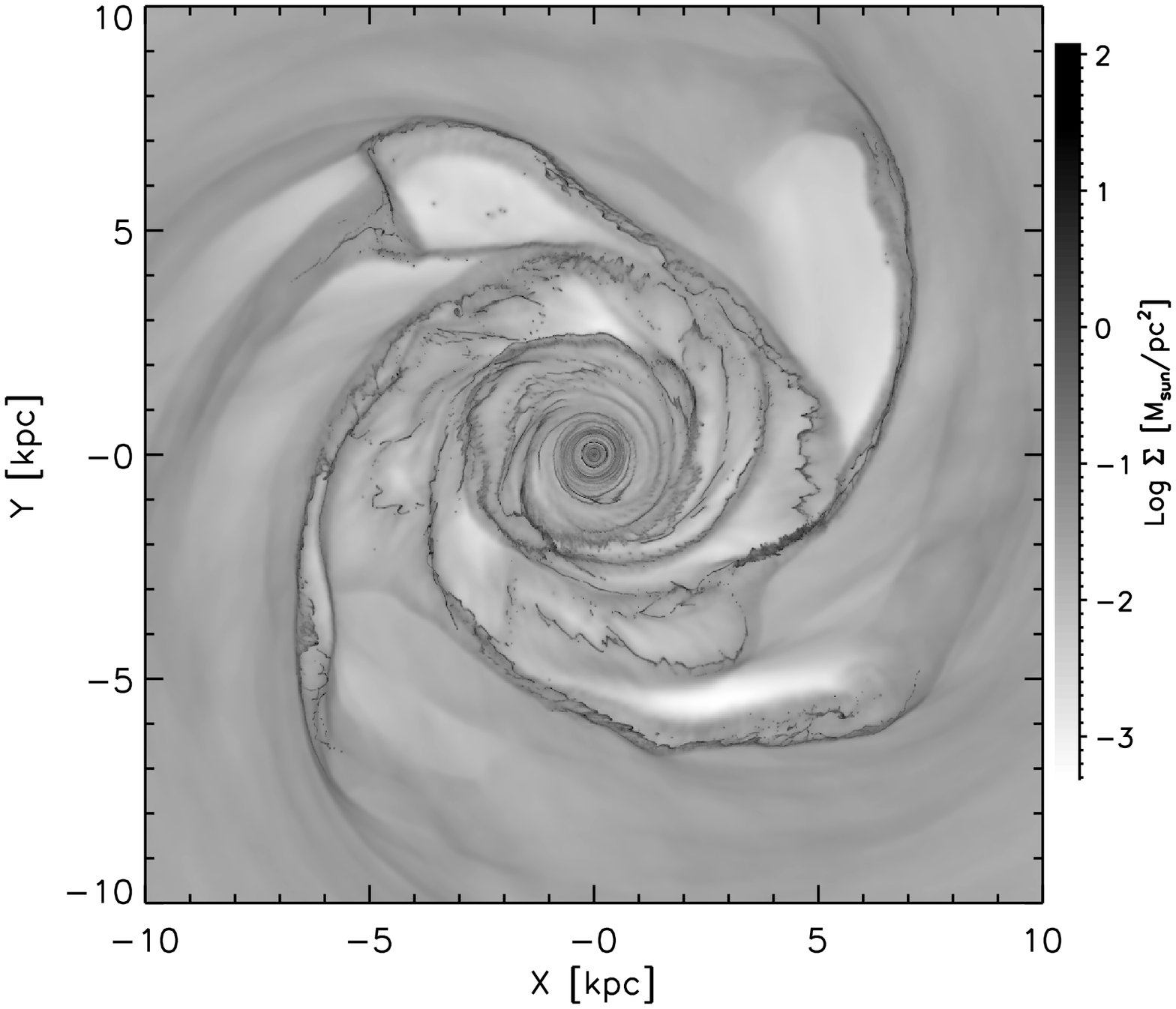}
  \includegraphics[width=36mm]{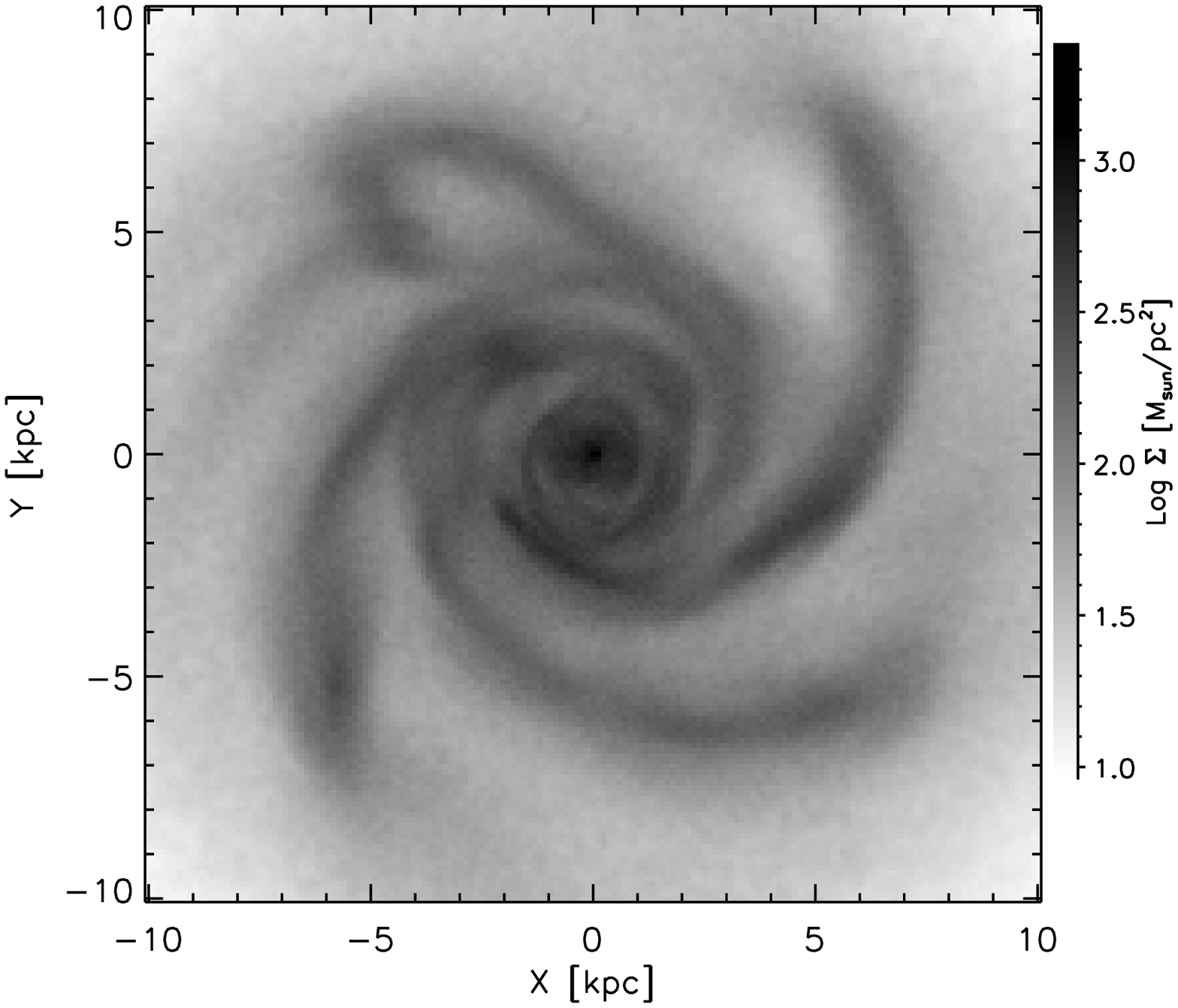}
  \includegraphics[width=36mm]{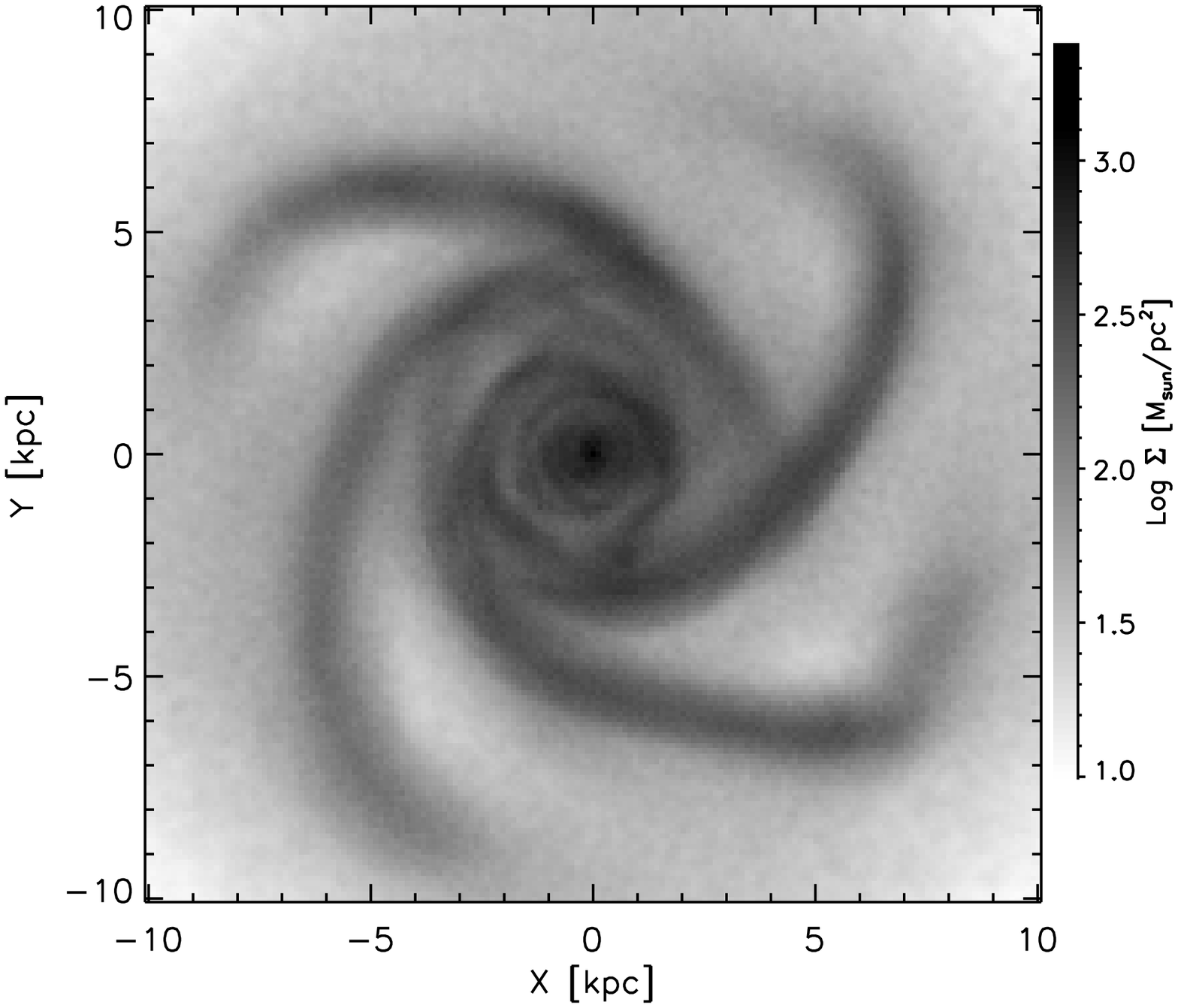}
  \includegraphics[width=36mm]{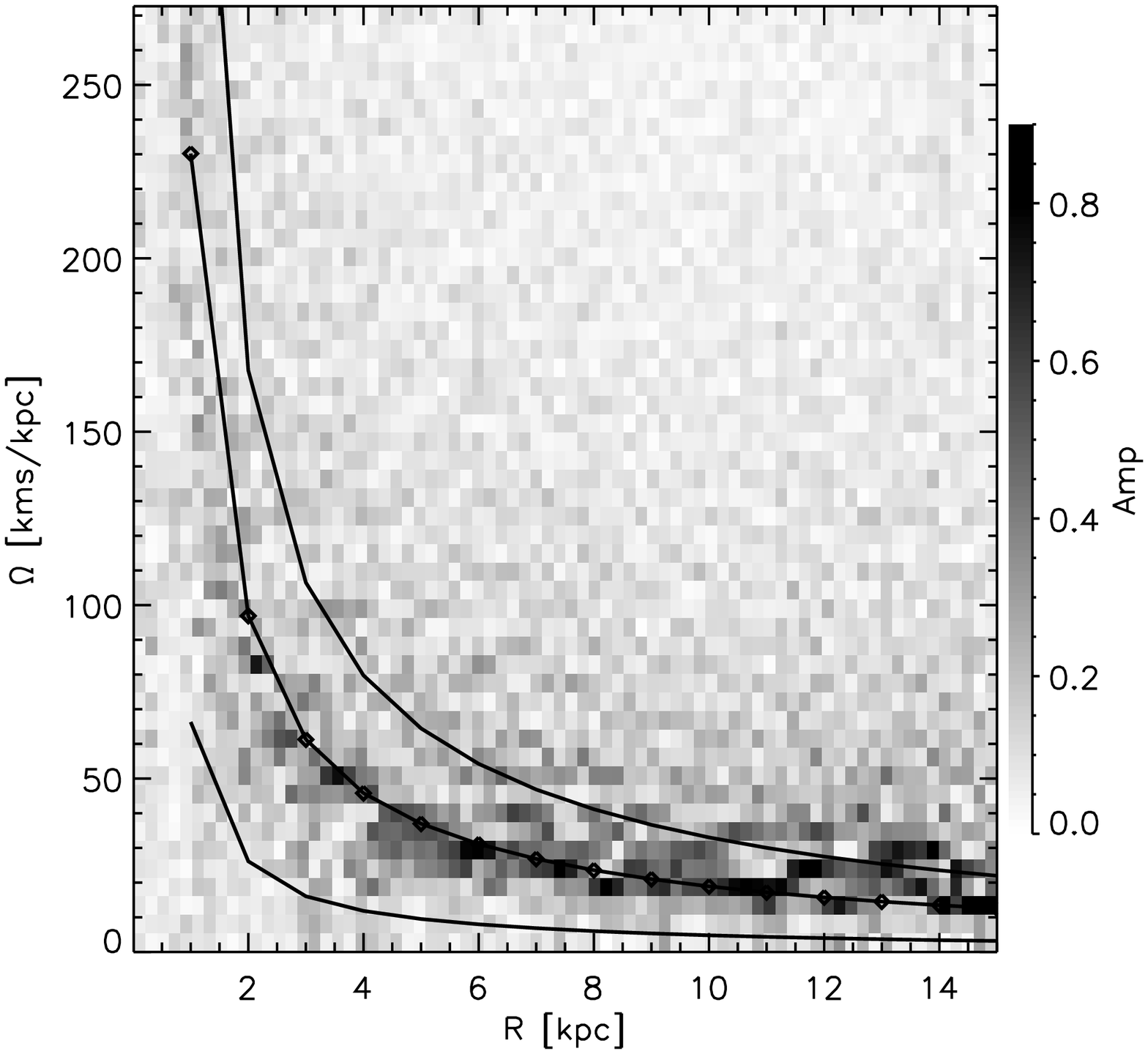}
  \includegraphics[width=36mm]{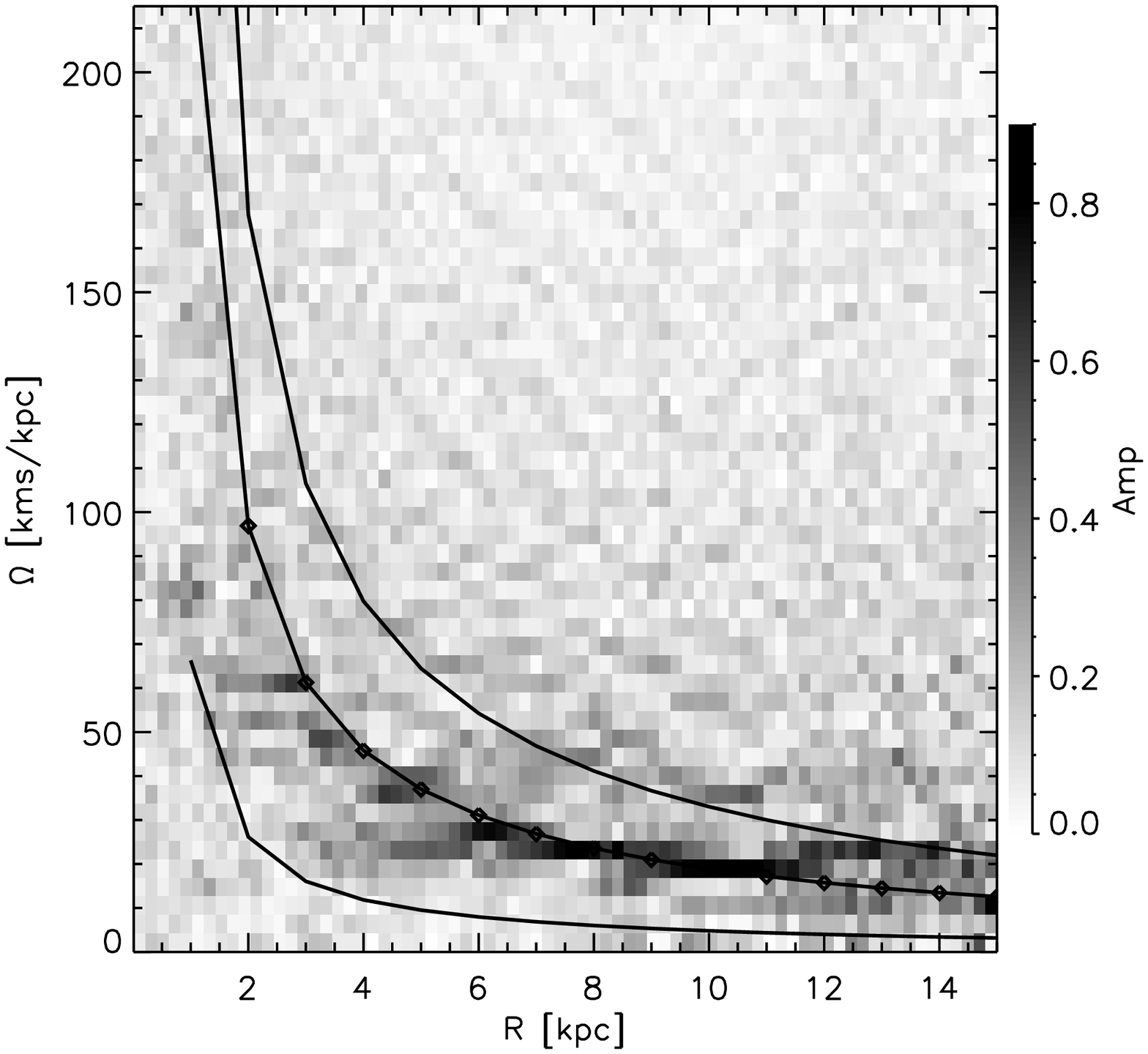}
  \includegraphics[width=36mm]{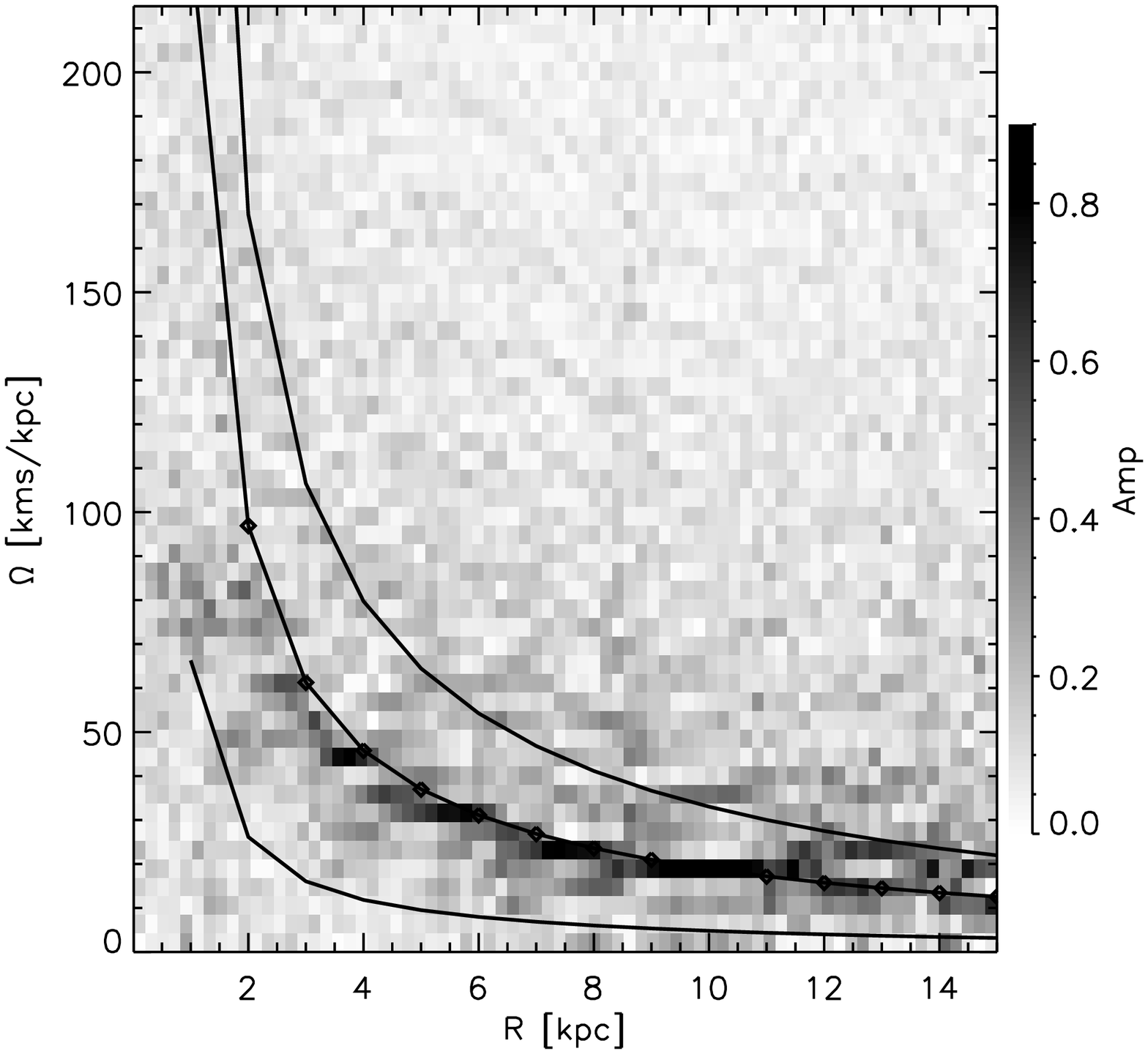}
  \caption{Left and center columns show the simulation with both components while the right column show the simulation with the stellar disk only, at $t = 200 \Myr$.}
   \label{dens}
 \end{center}
\end{figure*}

 The top row in Figure \ref{dens} shows the surface density distribution in both simulations. In a disk without gas, the spiral structure does not have substructure and the spiral pattern is stronger. The $m = 4$ mode is dominant in the simulation, so the pitch angle was calculated using the value of $p$ for the highest amplitude at  $t = 200\Myr$. We obtained pitch angle values of $25.2\grad$ and $22.8\grad$ for the stellar and gaseous components, respectively, in the first simulation, and $29.7\grad$ for the second one.

The bottom row in that same figure shows the spectrograms corresponding to the density distribution above. Top and bottom lines describe the inner and outer Lindblad's resonances while central line show the orbital frequency of the axisymmetric disk. On the gray scale, the highest values correspond to the spiral structure. It can be seen that, the dark parts match with the axisymmetric frequency in all the cases, suggesting that the spiral pattern co-rotates with the disk. 

\section{Conclusion}
We performed SPH+N-body simulations of galactic disk to explore differences in the structure of a stellar disk with and without a gaseous component. We found more substructure in a disk with both components. We decomposed both disks using Fourier analysis and calculated the pitch angle and the phase of the spiral structure. We found that the spiral structure co-rotates with the disk. 

\bibliography{paper}

\end{document}